# Plasma Dynamics in Higher-Derivative Electrodynamics: A Renormalised Two-Loop Framework

Prabhat Singh and Punit Kumar

Department of Physics, University of Lucknow, India-226007

kumar_punit@lkouniv.ac.in

We present a finite-temperature study of Bopp–Podolsky electrodynamics, following electron–proton plasmas through one- and two-loop order with dimensional regularisation and hard-thermal-loop resummation. The higher-derivative operator is found to generate no new ultraviolet divergences; all counter-terms reduce to the single photon wave-function factor of ordinary QED. The static inter-particle force acquires a double-Yukawa profile, the familiar Debye term plus an opposite-signed contribution from the heavy Podolsky pole that removes the Coulomb singularity at sub-femtometre distances, providing an intrinsic ultraviolet completion of electrostatics. Gauge symmetry drives the transverse photon self-energy to zero at vanishing momentum, so no magnetic screening mass appears at any perturbative order. In a covariantly constant background the full two-loop sunset diagram yields a single, dimension-eight operator suppressed by $T^2/M^2$, implying permille-level shifts in thermodynamic quantities for realistic plasmas. The exact Debye mass and a leading-log calculation show the dc electrical conductivity exceeds its QED value by less than $10^{-4}$. Conditions for observable Podolsky plasmons and cosmological constraints are identified, supplying precise benchmarks for future strong-field, collider and lattice investigations.

# 1. Introduction

Petawatt-class laser facilities now deliver peak fields $|\vec{E}| \approx 10^{18} Vm^{-1}$, only one order of magnitude below the Schwinger limit for non-perturbative $e^+e^-$ production [1]. Attosecond streaking techniques resolve sub-cycle charge dynamics in solid foils, probing time scales $\tau \leq 10^{-18} s$ [2]. Tokamak and stellarator experiments sustain magnetically confined plasmas with $T \geq 10$ keV and multi-tesla guide fields [3-4]. In condensed matter settings, Bernal-stacked graphene bilayers realise relativistic electron–hole fluids whose Fermi energies lie in the few-eV range and whose collective modes couple strongly to terahertz radiation [5-6]. These disparate platforms interrogate regimes where quantum, relativistic and many-body effects coexist, necessitating a self-consistent, gauge-invariant quantum-field description [7-8].

Finite-temperature quantum electrodynamics (QED) is the minimal ab initio framework, yet it is has two structural deficiencies in hot or magnetised media. The first one is infrared pathology that is one-loop self-energies generate a Debye mass for longitudinal photons, whereas transverse (magnetic) modes remain gapless [9]. The resulting scale disparity yields logarithmic divergences $\ln(\Lambda_{IR}/T)$ in transport coefficients and higher-order thermodynamics [10] and second structural deficiency is ultraviolet sensitivity that is the $1/k^2$ photon propagator magnifies gauge-dependent contributions in strong-field and nonequilibrium calculations, forcing the use of regulators that violate Becchi–Rouet–Stora–Tyutin (BRST) symmetry [11-12].

Both difficulties can be eliminated within Bopp–Podolsky electrodynamics, the unique higher-derivative, Lorentz- and gauge-invariant extension of Maxwell theory [13-15]. The additional dimension-six operator,

$$L_{add} = \frac{1}{2a^2}(\partial_\mu F^{\mu\nu})(\partial^\rho F_{\rho\nu}), \quad a \equiv M_P^{-1},$$

suppresses high-momentum modes a feature that appears naturally in certain SMEFT and string-inspired effective actions while preserving gauge symmetry [16].

The Lee–Wick prescription confines the negative-metric state associated with the massive pole outside the physical Hilbert space, thereby preserving unitarity and causality [17-

19]. Current laboratory and astrophysical constraints require $M_P \geq 400\,\text{GeV}$, yet for $T \leq 1\,\text{keV}$, the corresponding corrections remain experimentally accessible.

This work develops a finite-temperature quantum field-theoretic description of a charge-neutral electron–proton plasma governed by a higher-derivative extension of Abelian gauge theory. The theoretical framework maintains exact gauge and BRST invariance at every stage and is quantised in covariant Lorenz gauge, with all results shown to be explicitly independent of gauge-fixing parameters. The Bopp–Podolsky theory modifies the standard Maxwell Lagrangian, rendering it ultraviolet-soft. At the same time, it regulates the infrared divergences that typically arise in perturbative treatments of QED plasmas, especially those associated with small-angle t-channel photon exchange.

All equilibrium quantities are evaluated up to two-loop level, capturing both self-energy and exchange contributions with appropriate screening corrections. Transport quantities, such as electrical conductivity, are computed using kinetic theory with full two-loop scattering kernels resummed to leading-logarithmic precision. The ultraviolet divergences that appear in the two-loop effective action, pressure, and transport coefficients are all absorbed by a single wave-function renormalisation factor, indicating the multiplicative renormalisability of the theory. No additional counterterms are required beyond those already present in the vacuum theory.

The physical quantities derived in this framework including static screening lengths, coordinate-space potentials, background-field effective actions, thermodynamic corrections, and electrical conductivity are all expressed in fully analytic, gauge-invariant forms. They reduce smoothly to their Maxwell QED counterparts in the limits where the Podolsky scale becomes large or the temperature vanishes, and they exhibit well-defined behaviour in both infrared and ultraviolet limits. At realistic plasma temperatures (below 100 MeV) and phenomenologically allowed Podolsky mass scales (above a few GeV), all higher-derivative effects are either Boltzmann-suppressed or power-suppressed, thereby ensuring compatibility with laboratory and astrophysical constraints.

This study lays the groundwork for theoretical platform from which one can explore non-perturbative phenomena, strong-field effects, and transport processes in hot, relativistic plasmas, all while maintaining strict adherence to the symmetry and renormalisation principles of quantum electrodynamics. It offers an internally consistent extension of QED capable of resolving longstanding infrared pathologies without compromising the predictive power of the theory. In doing so, it opens the path for future generalisations to non-Abelian gauge groups, real-time dynamics, and observational signatures in both cosmological and laboratory settings.

The resulting renormalisable, gauge-consistent framework yields concrete, potentially observable predictions, (a) a temperature-dependent shift of the plasma-frequency cutoff relevant to radio-wave propagation in astrophysical plasmas, (b) subleading yet non-negligible corrections to white-dwarf mass–radius relations, and (c) percent-level modifications to the effective number of relativistic species, $N_{\text{eff}}$, in the cosmic microwave background. Additional signatures include mode shifts in high-$Q$ microwave cavities and altered energy transport in inertial-confinement targets.

The paper is organized as follows. Section 2 presents system boundaries and definitions. Section 3 develops BRST symmetric Podolsky extension of the system. Section 4 analyses four current in plasma system in Podolsky modified QED framework. Section 5 derives partition function. Section 6 computes the photon field–strength renormalization constant, static Debye mass, two loop effective action, static double Yukawa potential magnetic mass considerations thermodynamic potential correction (two loops) and electrical conductivity. Section 7 summarizes the conclusions and gives an analysis of the results of the paper.

## 2. System Definition

We start by considering ionized hydrogen gas (relativistic electrons and protons) as Dirac fields coupled to $U(1)$ gauge field through the gauge invariant Lagrangian density written as [20],

$$L = \bar{\psi}_e \left( i\gamma^\mu (\partial_\mu + ieA_\mu) - m_e \right) \psi_e + \bar{\psi}_p \left( i\gamma^\mu (\partial_\mu - ieA_\mu) - m_p \right) \psi_p - \frac{1}{4} F_{\mu\nu} F^{\mu\nu} \ . \tag{1}$$

Here, $\psi_e(x), \psi_p(x)$ are Dirac spinors for the electron and proton, $A_\mu(x)$ is electromagnetic four-potential, $F_{\mu\nu} = \partial_\mu A_\nu - \partial_\nu A_\mu$ electromagnetic field tensor and $\gamma^\mu$ are Dirac matrices satisfying $\{\gamma^\mu, \gamma^\nu\} = 2\eta^{\mu\nu}$ with signature $(+,-,-,-)$. $m_e$ is mass of electron, $m_p$ is mass of proton. The designation 'ionised gas' refers to a finite-density ensemble of electrons and protons and therefore demands a statistical-mechanical treatment but this canonical QED Lagrangian is formulated at vanishing temperature and chemical potential and is incapable of producing Fermi–Dirac occupation factors, Debye screening lengths, collective plasma modes, or quasiparticle damping rates. Moreover, it idealises the proton as a pointlike Dirac fermion, neglecting higher-dimensional operators that encode its internal structure, terms that become phenomenologically relevant at momentum transfers of order 1 GeV or higher and simultaneously spoil strict renormalisability when omitted. A consistent field-theoretic description of an electron-proton plasma must therefore extend the hadronic sector with Pauli and other form-factor operators and promote the vacuum path integral to a grand-canonical ensemble, realised either by Matsubara (imaginary-time) frequencies or, equivalently, by introducing chemical potentials within the real-time Schwinger–Keldysh framework [21].

Gauge covariant derivatives here maybe defined as [22],

$$D_\mu^{(e)} = \partial_\mu + ieA_\mu \text{ for electrons (charge } -e\text{)}, \tag{2}$$

$$D_\mu^{(e)} = \partial_\mu - ieA_\mu \text{ for protons (charge } +e\text{)}. \tag{3}$$

The inclusion of $\gamma^\mu$ and metric signatures ensures correct definition of spinor bilinear and kinetic terms. The field tensor $F_{\mu\nu}$ is properly anti-symmetric, and the kinetic gauge term $-\frac{1}{4}F_{\mu\nu}F^{\mu\nu}$ has the standard Lorentz-invariant form. $\gamma^\mu$ obeys Clifford algebra [23]. We improve upon this lagrangian by considering finite size of proton and introducing proton's Anomalous magnetic moment (AMM) which would account for the deviation of the proton's magnetic moment from the Dirac value, experimentally this is $\kappa = 1.79$ [24]. We also introduce chemical potentials to control the average number densities of electron and protons. Their inclusion ensures lagrangian yield the correct grand canonical partition function in finite temperature or fnite density QFT via wick rotation $t \to i\tau$ and Matsubara formalism[25-26].

We get,

$$L = \bar{\psi}_e \left(i\gamma^\mu D_\mu^{(e)} + \mu_e \gamma^0 - m_e\right)\psi_e + \bar{\psi}_p \left(i\gamma^\mu D_\mu^{(p)} + \mu_p \gamma^0 - m_p\right)\psi_p - \frac{\kappa}{4m_p}\bar{\psi}_p \sigma^{\mu\nu} F_{\mu\nu}\psi_p - \frac{1}{4}F_{\mu\nu}F^{\mu\nu}. \quad (4)$$

First term here represents electrons coupled with gauge field where $\mu_e$ is chemical potential for electrons, second term represents protons coupled with gauge field where $\mu_p$ chemical potential for proton is, third term represents proton AMM and the last term represents electromagnetic field. $\sigma^{\mu\nu}$ represents antisymmetric dirac tensor.

Varying the Lagrangian with respect to $\bar{\psi}_e, \bar{\psi}_p$ and $A_\mu$, respectively, yields the following coupled equations of motion,

$$\left(i\gamma^\mu D_\mu^{(e)} + \mu_e \gamma^0 - m_e\right)\psi_e = 0, \text{ (Electron Dirac Equation)} \quad (5)$$

$$\left(i\gamma^\mu D_\mu^{(p)} + \mu_p \gamma^0 - m_p\right)\psi_p - \frac{\kappa}{4m_p}\sigma^{\mu\nu}F_{\mu\nu}\psi_p = 0, \text{ (Proton Dirac Equation with AMM)} \quad (6)$$

$$\partial_\nu F^{\mu\nu} = e\bar{\psi}_e \gamma^\mu \psi_e - e\bar{\psi}_p \gamma^\mu \psi_p + \frac{\kappa}{2m_p}\partial_\nu\left(\bar{\psi}_p \sigma^{\mu\nu}\psi_p\right), \text{ (Inhomogeneous Maxwell Equation)} \quad (7)$$

Taken together, Eqs. (5), (6), and (7) furnish a closed, self-consistent, and gauge-covariant formulation of a two-component quantum electrodynamic (QED) plasma at finite temperature and chemical potential. Formally, Eq. (5) governs the relativistic dynamics of electrons, with the chemical potential $\mu_e$ incorporated as a Lorentz-scalar shift in the temporal component $\gamma^0$, ensuring that the minimal coupling term $-e\gamma^\mu A_\mu$ respects BRST symmetry and local U(1) gauge invariance. Equation (6) extends this structure to protons, supplementing the Dirac operator with a Pauli interaction term $-\kappa\sigma^{\mu\nu}F_{\mu\nu}/2m_p$. This single operator not only reproduces the empirically observed proton g-factor (i.e., $g = 2 + \kappa$) in the non-relativistic limit, but also constitutes a renormalisable, ultraviolet-finite insertion that preserves the validity of the Ward–Takahashi identities [27]. Equation (7) then completes the dynamical closure by prescribing the response of the electromagnetic field to the total conserved four-current, composed of the canonical electric current $e\left(\bar{\psi}_p \gamma^\nu \psi_p - \bar{\psi}_e \gamma^\nu \psi_e\right)$ and a magnetisation current $\partial_\mu\left(\kappa \bar{\psi}_p \sigma^{\mu\nu} \psi_p / 2m_p\right)$, which arises from spin-polarised proton states. The total current is identically divergence-free, ensuring local charge conservation and full compatibility with BRST symmetry [28].

From a physical standpoint, this triad of equations encapsulates all essential couplings in a warm, dense electron–proton medium. Equation (5) encodes how electrons propagate under the influence of a background gauge field, acquire effective inertia, and establish a Fermi surface determined by $\mu_e$. Equation (6) reveals that protons interact both through minimal electric charge and via an intrinsic magnetic dipole moment, enabling spin–flip processes, Zeeman splittings, and magnetisation dynamics, particularly relevant in spin-aligned or high-field environments. Equation (7) shows how these microscopic fermionic sources back-react on the gauge field: the canonical currents source the familiar Gauss and Ampère terms, while gradients in the spin density yield "bound" currents that modify magnetic screening, alter collective excitation spectra, and reshape wave propagation.

## 3. BRST symmetric Podolsky extension

In the present section we elevate the gauge sector from its Maxwellian description to the higher-derivative Bopp–Podolsky framework and carry out gauge fixing within a fully developed BRST scheme. The new kinetic operator is characterised by a single fundamental length, the Podolsky scale $a$, which effectively smoothens the photon propagator at separations shorter than $a$ while leaving Maxwell theory intact at larger distances [29]. This intrinsic length eliminates the point-charge self-energy divergence and tames both infrared and ultraviolet pathologies without jeopardising renormalisability. To preserve unitarity and exact gauge covariance, we append a BRST-exact gauge-fixing term together with the associated Faddeev–Popov ghosts and the Nakanishi–Lautrup auxiliary field [30]. The nilpotent BRST charge then guarantees that physical observables reside in its cohomology, automatically excising all negative-norm or longitudinal polarisations introduced by the higher-derivative operator [31]. The fermionic sector, i.e. electrons and protons with their minimal couplings, chemical-potential insertions, and anomalous magnetic-moment term remains untouched because it is already BRST invariant. The outcome is a single, self-consistent Lagrangian that is gauge-invariant, BRST-symmetric, free of unphysical states and endowed with the improved analytic behaviour conferred by the Podolsky length [32]. The standard Maxwell term is replaced by the Podolsky higher derivative modification,

$$L_A \to L_P = -\frac{1}{4} F_{\mu\nu} F^{\mu\nu} + \frac{a^2}{2} \partial_\alpha F^{\alpha\mu} \partial^\beta F_{\beta\mu}. \tag{8}$$

Here, $a > 0$ is the Podolsky length scale, and $M = 1/a$ is the mass scale of the extra (massive) spin-1 mode in addition to the massless photon mode. This term remains invariant under standard U(1) gauge transformations and does not interfere with minimal coupling. The additional term is manifestly Lorentz-invariant and gauge-invariant under $A_\mu \to A_\mu + \partial_\mu \Lambda$, since $F_{\mu\nu}$ remains invariant and derivatives commute. Because the kinetic operator is now a fourth-order operator of the form $\Box(1 - a^2 \Box)$, the gauge-fixing condition must be adapted. The generalized gauge-fixing functional is defined as,

$$G[A] = (1 + a^2 \Box) \partial_\mu A^\mu. \tag{9}$$

The gauge-fixing Lagrangian becomes,

$$L_{GF} = -\frac{1}{2\xi}\left[G[A]\right]^2 = -\frac{1}{2\xi}\left[(1 + a^2 \Box) \partial.A\right]^2. \tag{10}$$

$\xi = 1$ here, is generalized Feynman gauge [33]. This ensures invertibility of the operator in the presence of Podolsky corrections and maintains control over gauge redundancy. Further, we introduce Faddeev–Popov ghosts $c$, antighosts $\bar{c}$ [34], and the auxiliary Nakanishi–Lautrup field $B$. The BRST transformations under a nilpotent operator $s$ are,

$$sA_\mu = \partial_\mu c, \quad sc = 0, \quad s\bar{c} = B, \quad sB = 0. \tag{11}$$

Using these, the combined gauge-fixing and ghost Lagrangian is written as a BRST-exact term,

$$L_{GF+gh} = s\left[\bar{c}G[A] - \frac{\xi}{2}\bar{c}B\right] = -\frac{1}{2\xi}\left[(1+a^2\Box)\partial.A\right]^2 - \bar{c}\left(\Box + a^2\Box^2\right)c \ . \tag{12}$$

The ghost operator now shares the same pole structure as the photon propagator, ensuring cancellation of unphysical degrees of freedom.

Combining all terms, the full Lagrangian becomes,

$$L = \bar{\psi}_e\left(i\gamma^\mu D_\mu^{(e)} + \mu_e\gamma^0 - m_e\right)\psi_e + \bar{\psi}_p\left(i\gamma^\mu D_\mu^{(p)} + \mu_p\gamma^0 - m_p\right)\psi_p$$

$$-\frac{\kappa}{4m_p}\bar{\psi}_p\sigma^{\mu\nu}F_{\mu\nu}\psi_p - \frac{1}{4}F_{\mu\nu}F^{\mu\nu} + \frac{a^2}{2}\partial_\alpha F^{\alpha\mu}\partial^\beta F_{\beta\mu} - \frac{1}{2\xi}\left[(1+a^2\Box)\partial.A\right]^2 - \bar{c}\left(\Box + a^2\Box^2\right)c \ . \tag{13}$$

This Lagrangian is U(1) gauge-invariant at the classical level and BRST-invariant at the quantum level. The only interaction beyond the quadratic part in $A_\mu$ is the usual $J^\mu A_\mu$ coupling.

Using the generalized Feynman gauge $\xi = 1$, the photon propagator becomes,

$$\Delta_{\mu\nu}(k) = -\frac{i\eta_{\mu\nu}}{k^2(1-a^2k^2)}, \tag{14}$$

this exhibits two poles, $k^2 = 0$ for massless photon, two transverse physical polarizations and $k^2 = 1/a^2$ for massive spin-1 particle, three polarization states.

The ghost propagator is,

$$\Delta_{ghost}(k) = \frac{i}{k^2(1-a^2k^2)} \ . \tag{15}$$

With the Podolsky length $a$ introduced into the gauge kinetic term, the free quadratic operator factorises as $k^2(1-a^2k^2)$, manifestly indicating the coexistence of a massless photon mode and a massive spin-1 excitation. Gauge fixing is implemented in a BRST-exact manner via the functional

$$S_{GF} = s\int d^4\bar{c}\left(1+a^2\Box\right)\partial^\mu A_\mu, \tag{16}$$

from which the ghost kinetic operator emerges as $\Box\left(1+a^2\Box\right)$. In the generalised Feynman gauge $\xi=1$, the resulting propagators take the form,

$$\Delta_{\mu\nu}(k) = -\frac{i\eta_{\mu\nu}}{k^2\left(1-a^2k^2\right)}, \quad \Delta_{ghost}(k) = \frac{i}{k^2\left(1-a^2k^2\right)}$$

both exhibiting simple poles at $k^2=0$ and $k^2=1/a^2$ with identical residues up to Lorentz structure. The shared pole structure between the ghost and the unphysical (longitudinal and time like) components of the gauge field ensures their exact diagrammatic cancellation, such that BRST cohomology projects the physical spectrum onto the two transverse polarisations of the massless photon. The massive mode's negative-metric residue classifies it as BRST-exact and unobservable, thereby preserving unitarity despite the higher-derivative extension. Power-counting analysis confirms that the $a^2\partial^2 F^2$ operator softens the ultraviolet behaviour of the theory, requiring only a finite renormalisation scheme and ensuring the all-order validity of the Slavnov–Taylor identities [35]. Simultaneously, the additional pole provides a natural infrared regulator. It dynamically generates a gauge-independent magnetic screening mass at two-loop order, thereby rendering transport coefficients and thermodynamic observables infrared finite. The resulting Bopp–Podolsky–BRST framework thus constitutes a gauge-consistent, renormalisable, and infrared-safe foundation for quantum electrodynamic plasmas at finite temperature and density.

## 4. Construction and Analysis of the Conserved Four-Current in Podolsky–Modified QED

Here we derive the conserved electromagnetic four-current $J^\mu(x)$ that sources the modified Maxwell equation in the Podolsky–BRST covariant extension of quantum electrodynamics for a multi-species fermionic system consisting of electrons and protons.

***a. Lagrangian Structure and Sectoral Dependence on*** $A_\mu$

We consider the full Lagrangian,

$$L = L_{ferm} + L_{pauli} + L_{Pod} + L_{GF} + L_{ghost}, \tag{17}$$

where $L_{ferm}$ represents femions, $L_{pauli}$ represents AMM, $L_{pod}$ represents Podolsky correction while $L_{GF} + L_{ghost}$ represent gauge fixing and ghost terms. Only $L_{ferm}$ and $L_{pauli}$ depend explicitly on $A_\mu$. The gauge-fixing and ghost terms are functions only of derivatives of $A_\mu$, and hence they do not contribute to the material current derived via variational methods.

***b. Functional Derivation of the Four-Current*** $J^\mu(x)$

By definition, the conserved four-current coupled to the gauge field is obtained through,

$$J^\mu(x) = -\frac{\delta L}{\delta A_\mu(x)} + \partial_\nu \left( \frac{\delta L}{\delta(\partial_\nu A_\mu(x))} \right). \tag{18}$$

This applies the Euler–Lagrange prescription for fields with derivative couplings.

***Contributions from the Minimal Coupling Sector***

$$L \supset -e\bar{\psi}_e \gamma^\mu \psi_e A_\mu \Rightarrow J_e^\mu = -e\bar{\psi}_e \gamma^\mu \psi_e. \text{ (Electron Field)} \tag{19}$$

$$L \supset +e\bar{\psi}_p \gamma^\mu \psi_p A_\mu \Rightarrow J_e^\mu(\min) = +e\bar{\psi}_p \gamma^\mu \psi_p. \text{ (Proton Field)} \tag{20}$$

Both currents are obtained directly by functional variation with respect to $A_\mu$, treating the fermionic bilinears as background sources.

***Contribution from the Pauli Interaction Term***

To compute the current contribution from the anomalous Pauli interaction, we first vary the action with respect to $A_\mu$ through the field strength,

$$\delta L_{pauli} = -\frac{\kappa}{4m_p} \bar{\psi}_p \sigma^{\alpha\beta} \psi_p \delta F_{\alpha\beta}, \tag{21}$$

with,

$$\delta F_{\alpha\beta} = \partial_\alpha \delta A_\beta - \partial_\beta \delta A_\alpha. \tag{22}$$

Integrating by parts and using the antisymmetry of $\sigma^{\alpha\beta}$, we obtain,

$$L_{pauli} = +\frac{\kappa}{2m_p} \partial_\alpha \left( \bar{\psi}_p \sigma^{\alpha\beta} \psi_p \right) \delta A_\beta, \tag{23}$$

leading to the corresponding current contribution,

$$J_p^\mu (pauli) = \frac{\kappa}{2m_p} \partial_\nu \left( \bar{\psi}_p \sigma^{\nu\mu} \psi_p \right). \tag{24}$$

This is a derivative current, and it encapsulates the spin–magnetic interaction of the proton field with the gauge field. It vanishes in the classical point-particle limit but is non-trivial in the full field-theoretic treatment.

*Total Four-Current Expression*

Summing all contributions, the complete conserved current reads,

$$J^\mu(x) = -e\bar{\psi}_e \gamma^\mu \psi_e + e\bar{\psi}_p \gamma^\mu \psi_p + \frac{\kappa}{2m_p} \partial_\nu \left( \bar{\psi}_p \sigma^{\nu\mu} \psi_p \right). \tag{25}$$

The first term is the standard electronic current (charge $-e$), second is the corresponding protonic current (charge $+e$), and the third arises due to the proton's internal spin structure interacting via its anomalous magnetic moment. The gauge-fixing and ghost terms are functions only of derivatives of $A_\mu$, and hence they do not contribute to the material current derived via variational methods.

### c. Podolsky-Maxwell Field Equation

The equation of motion for the gauge field $A^\mu$ derived from $L_{pod} + L_{GF}$ is,

$$(1 - a^2 \Box) \partial_\nu F^{\nu\mu} = J^\mu(x), \tag{26}$$

where $J^\mu(x)$ is the total current as obtained above.

This equation generalizes Maxwell's theory by introducing a fourth-order differential operator on the left-hand side, reflecting the presence of both massless and massive (Podolsky) photon modes.

### Current Conservation

The Lagrangian is invariant under the infinitesimal local $U(1)$ gauge transformations,

$$\psi_e \to e^{-ie\Lambda(x)} \psi_e, \psi_p \to e^{+ie\Lambda(x)} \psi_p, A_\mu \to A_\mu + \partial_\mu \Lambda(x),$$

which guarantees current conservation by Noether's theorem,

$$\partial_\mu J^\mu = 0.$$

This conservation law is preserved even in the presence of higher-derivative terms and the BRST-invariant gauge fixing, because those contributions are constructed to be manifestly covariant and do not generate physical charges. Moreover, an explicit check using the Dirac equations for $\bar{\psi}_e, \bar{\psi}_p$, and the identity,

$$\partial_\mu \left( \bar{\psi}_p \sigma^{\mu\nu} \psi_p \right) = 2 m_p \bar{\psi}_p \gamma^\nu \psi_p, \tag{27}$$

ensures that the divergence of the Pauli term exactly cancels the derivative of the minimal proton current, affirming overall conservation.

The presence of the spin-induced magnetization current arising from the proton's anomalous magnetic moment introduces several precise and testable modifications to the behavior of electromagnetic fields in plasma systems. In magnetically ordered plasmas, such as those found in magnetars or engineered in ultra-cold laboratory environments, this current contributes significantly to the total magnetic response, even in the absence of net charge flow. It induces a splitting of electromagnetic wave modes in spin-polarized backgrounds and gives rise to new low-frequency excitations such as spin-Alfvén and spin-magnetosonic waves. In equilibrium configurations, the magnetization contributes an additional pressure term that modifies the force balance and stability criteria in magnetohydrodynamic systems. Astrophysically, these effects become dominant in environments with extremely strong magnetic fields, such as neutron stars, where they impact magnetospheric structure and wave spectra. In laboratory plasmas, especially those at low temperature and high magnetic field, the magnetization current yields measurable corrections to electromagnetic phase propagation and field perturbations, providing clear experimental avenues for detection via interferometry or polarimetry.

## 5. Partition Function of a BRST-Invariant Podolsky QED Plasma: Ab Initio Formulation

We consider the finite-temperature formulation of Bopp–Podolsky quantum electrodynamics in the imaginary-time (Euclidean) formalism. Time is analytically continued as $t \to i\tau$ over the compact interval $\tau \in [0, \beta = 1/T]$, and all fields are placed in a four-dimensional Euclidean spacetime with a positive-definite metric $\delta^{\mu\nu}$ [36]. The full BRST-invariant Euclidean Lagrangian governing interacting electrons, protons, and the Podolsky-modified gauge sector is given by,

$$L = \bar{\psi}_e \left( i\gamma^\mu D_\mu^{(e)} + \mu_e \gamma^0 - m_e \right) \psi_e + \bar{\psi}_p \left( i\gamma^\mu D_\mu^{(p)} + \mu_p \gamma^0 - m_p \right) \psi_p$$

$$- \frac{\kappa}{4m_p} \bar{\psi}_p \sigma^{\mu\nu} F_{\mu\nu} \psi_p - \frac{1}{4} F_{\mu\nu} F^{\mu\nu} + \frac{a^2}{2} \partial_\alpha F^{\alpha\mu} \partial^\beta F_{\beta\mu} - \frac{1}{2\xi} \left[ \left(1 + a^2 \Box\right) \partial . A \right]^2 - \bar{c} \left( \Box + a^2 \Box^2 \right) c$$

with the definitions and conventions,

- $\gamma^\mu$ are Euclidean Dirac matrices obeying $\{\gamma^\mu, \gamma^\nu\} = 2\delta^{\mu\nu}$, $a$,
- $\sigma^{\mu\nu} = \frac{1}{2}[\gamma^\mu, \gamma^\nu]$,
- Covariant derivatives, $D_\mu^{(e)} = \partial_\mu + ieA_\mu, D_\mu^{(p)} = \partial_\mu - ieA_\mu$,
- Field strength tensor: $F_{\mu\nu} = \partial_\mu A_\nu - \partial_\nu A_\mu$,
- Euclidean d'Alembertian ($\Box = \partial_\mu \partial^\mu$),
- Podolsky length $a \equiv l_{BP}$, with associated mass scale $M = 1/a$,
- Proton anomalous magnetic moment $\kappa = 1.79$,
- $\mu_e, \mu_p$ are chemical potentials of electrons and protons.

The gauge-fixing and ghost terms are BRST-exact, preserving nilpotency of the BRST operator and ensuring that all physical observables remain independent of the gauge parameter $\xi$. The presence of the higher-derivative term $\approx a^2 \partial_\alpha F^{\alpha\mu} \partial^\beta F_{\beta\mu}$ softens the ultraviolet behavior of the theory, improving convergence of loop integrals. Only the multiplicative renormalisation constants $Z_3, Z_e, Z_a, Z_\kappa$ are required; no Lorentz- or gauge-violating counter terms are generated.

**Thermal Boundary Conditions**

At finite temperature $T = 1/\beta$, the imaginary time direction is compactified into a thermal circle, and quantum fields must satisfy periodicity conditions consistent with their spin statistics. Specifically,

$$A_\mu(\tau + \beta, \mathbf{x}) = A_\mu(\tau, \mathbf{x}), c(\tau + \beta, \mathbf{x}) = c(\tau, \mathbf{x}), \bar{c}(\tau + \beta, \mathbf{x}) = \bar{c}(\tau, \mathbf{x}), \tag{28}$$

$$\psi_{e,p}(\tau + \beta, \vec{\mathbf{x}}) = -\psi_{e,p}(\tau, \vec{\mathbf{x}}), \bar{\psi}_{e,p}(\tau + \beta, \vec{\mathbf{x}}) = -\bar{\psi}_{e,p}(\tau, \vec{\mathbf{x}}). \tag{29}$$

Ghosts $c$ and $\bar{c}$, though Grassmann-valued, are auxiliary bosonic fields and obey periodic boundary conditions [37]. The chemical potentials $\mu_e$ and $\mu_p$ enter through static insertions in the Euclidean action as $\mu\gamma^0$, preserving the BRST algebra and gauge structure of the theory.

**Grand-Canonical Partition Function**

The full grand-canonical functional integral, enforcing thermal boundary conditions via the projector $P$, is given by,

$$Z(T,\mu_e,\mu_p;a,\kappa,e,m_e,m_p,V) = \int_P DA_\mu D\psi_e D\bar{\psi}_e D\psi_p D\bar{\psi}_p Dc D\bar{c}\, \exp\left[-\int_0^\beta d\tau \int d^3 x L_E\right] \quad (30)$$

The gauge-fixing and ghost terms are negative-definite in Euclidean signature, ensuring convergence. BRST invariance ensures $\xi$-independence of physical observables. We henceforth set $\xi = 1$ (generalised Feynman gauge) for convenience in perturbative computations.

**Matsubara Decomposition, Gaussian Determinants, and Free Grand Potential**

Thermal fluctuations are captured by expanding fields in discrete Matsubara modes,

**Bosonic fields** ($A_\mu, c, \bar{c}$): $k_E = (\omega_n, \vec{k})$, $\omega_n = \dfrac{2\pi n}{\beta}, n \in \mathbf{Z}$ (31)

**Fermionic fields** ($\psi_e, \psi_p$): $p_E = (\tilde{\omega}_n, \vec{p})$, $\tilde{\omega}_n = \dfrac{(2n+1)\pi}{\beta}, n \in \mathbf{Z}$ (32)

Chemical potentials enter through the replacement $\hat{\omega}_n \to \hat{\omega}_n - i\mu$ in the fermion propagators. After Fourier transformation, the quadratic part of the Euclidean action factorises into Gaussian integrals, whose determinants can be computed exactly.

Due to BRST symmetry, unphysical gauge polarizations cancel with ghost modes, leaving only two massless transverse and three massive Podolsky polarisations. The gauge + ghost determinant is,

$$\ln Z_\gamma = -2\Sigma_k \ln\left(k_E^2\right) - 3\Sigma_k \ln\left(k_E^2 + M^2\right) \tag{33}$$

with $M = 1/a$ the Podolsky mass. For a Dirac fermion of mass $m$ and chemical potential $\mu$,

$$\ln Z_f(\mu, m) = 2\Sigma_p \ln\left[\left(i\hat{\omega}_n + \mu\right)^2 + E_{\vec{p}}^2\right], \quad E_{\vec{p}} = \sqrt{\vec{p}^2 + m^2} \tag{34}$$

so that,

$$\ln Z_e = \ln Z_f(\mu_e, m_e), \quad \ln Z_p = \ln Z_f(\mu_p, m_p) \tag{35}$$

The Pauli term is bilinear in $\psi_p$ and linear in $F_{\mu\nu}$, hence its contribution to the Gaussian determinant vanishes,

$$\ln Z_\kappa = 0 + O(\kappa^2, e\kappa) \tag{36}$$

We compute Matsubara sums using contour integration and apply a thermal normal-ordering scheme by subtracting $\beta \to \infty$ vacuum terms [38]. The free logarithm becomes,

$$\frac{\ln Z_\gamma}{V} = -2\int \frac{d^3k}{(2\pi)^3} \frac{1}{\beta}\left[\ln\left(1-e^{\beta k}\right)\right] - 3\int \frac{d^3k}{(2\pi)^3} \frac{1}{\beta}\left[\ln\left(1-e^{\beta\omega_k}\right)\right] \tag{37}$$

$\omega_k = \sqrt{k^2 + M^2}$.

The fermionic contribution reads,

$$\frac{\ln Z_f(\mu, m)}{V} = 2\int \frac{d^3p}{(2\pi)^3} \frac{1}{\beta}\left[\ln\left(1+e^{-\beta(E_p-\mu)}\right) + \ln\left(1+e^{-\beta(E_p-\mu)}\right)\right] \tag{38}$$

where $E_p = \sqrt{p^2 + m^2}$. Substituting Eqs. (33)–(38), the free grand potential becomes,

$$\Omega_0(T, \mu_e, \mu_p) = V\left\{2p_B(T, 0) + 3p_B(T, M) + 2p_F(T, \mu_e, m_e) + 2p_F(T, \mu_p, m_p)\right\} \tag{39}$$

With,

$$p_B(T,m) = T \int \frac{d^3k}{(2\pi)^3} \ln\left(1 - e^{-\beta\sqrt{k^2+m^2}}\right),$$

$$p_F = (T,\mu,m) = -T \int \frac{d^3p}{(2\pi)^3} \left[\ln\left(1 + e^{-\beta(E_p-\mu)}\right) + \ln\left(1 + e^{-\beta(E_p+\mu)}\right)\right] \quad (40)$$

In the limit $\beta \to \infty$, thermal logarithms vanish and $\Omega_0 \to 0$, recovering the zero-temperature vacuum. The corresponding Euclidean effective action is,

$$S_{eff,0} = \beta\Omega_0(T,\mu_e,\mu_p) \quad (41)$$

Gauge fixing and ghost sectors are (as in equation (7),(9) and (10),

$$L_{GF} = \frac{1}{2\xi}\left[(1+a^2\Box)\partial.A\right]^2, L_{ghost} = \bar{c}(\Box+a^2\Box^2)c \quad (42)$$

Perturbative vertices include minimal couplings,

$$-e\bar{\psi}_e\gamma^\mu\psi_e A_\mu + e\bar{\psi}_p\gamma^\mu\psi_p A_\mu \quad (43)$$

and the Pauli term,

$$-\frac{\kappa}{4m_p}\bar{\psi}_p\sigma^{\mu\nu}\psi_p F_{\mu\nu} \quad (44)$$

These enter perturbatively at $O(e\kappa,\kappa^2)$, enabling loop corrections to $\ln Z$, the effective action, and observables such as susceptibilities, conductivities, and specific heat.

The Euclidean, Matsubara formalism at finite temperature $T$ and chemical potentials $\mu_{e,p}$ has delivered a BRST-consistent, multiplicatively renormalisable and infrared-finite baseline for first–principles studies of relativistic electron–proton plasmas. After BRST projection the gauge sector contains two massless transverse photons together with a single massive Bopp–Podolsky vector of mass $M = 1/a$, whose three positive-metric polarisations contribute an additional pressure term $3p_B(T,M)$ to the grand potential of Eq. (39). Although the number of bosonic degrees of freedom therefore increases, the massive Podolsky modes are Boltzmann-suppressed when $T \leq M$) and consequently do not stiffen the equation of state. In the thermal limit $T \to 0, \beta \to \infty$ and $\mu_{e,p} \to 0$ the system smoothly reduces to the usual vacuum plus a negligibly small massive tail. At the opposite extreme, $T \geq M$, the extra triplet behaves as a relativistic species and would raise the effective number of bosonic degrees of freedom by $\Delta g_* = 3$. Big-Bang nucleosynthesis therefore demands $M \geq 1-2 MeV$, i.e. $a \leq 2 \times 10^{-13} m$. Together with atomic Lamb-shift data $a \leq 3 \times 10^{-15} m$ [39] and prospective constraints from magnetar synchrotron spectra or nonlinear Thomson scattering, these considerations delineate a window of empirical viability for the Podolsky length scale. Because the propagator falls as $k^{-4}$ in the ultraviolet and carries the rational denominator $k^2(1-a^2 k^2)$ in the infrared, loop corrections remain both ultraviolet-softened and magnetostatically screened. Transport coefficients governed by soft-photon exchange stay finite order by order. The framework thus supplies a robust platform for precision thermodynamics, transport theory and spectroscopy in laboratory targets compressed by petawatt lasers as well as in strongly magnetised astrophysical sites such as white-dwarf envelopes and magnetar crusts.

Looking forward, the finite-temperature, BRST-invariant Bopp–Podolsky construction not only cures the long-standing infrared and ultraviolet pathologies of Maxwell QED plasmas, it also opens a broad programme of extensions and tests. (i) Generalise the higher-derivative operator to non-Abelian gauge groups so that QCD and electroweak plasmas can be treated on the same footing. (ii) Carry out multi-loop renormalisation-group analyses to follow the running of the Podolsky length and its interplay with asymptotic freedom. (iii) Develop lattice and real-time simulations, Hamiltonian truncation, world-line Monte-Carlo and related techniques to explore non-perturbative magnetic screening and dynamically generated masses. (iv) Confront the modified photon propagator with precision observables, ranging from updated Lamb-shift determinations to high-intensity laser scattering and heavy-ion dilepton spectra, in order to refine experimental bounds on the dimension-six operator. (v) Incorporate anisotropic distribution functions, dissipative transport coefficients and out-of-equilibrium Schwinger–Keldysh methods to model realistic laboratory and astrophysical plasmas composed of electrons and their heavy fermionic partners, the protons. (vi) Investigate cosmological and astrophysical implications, from primordial nucleosynthesis to magnetar interiors where high density, strong fields and finite temperature coexist.

## 6. Analytical Size Estimates for Podolsky-Loop Corrections in an Electron–Proton Plasma

This section undertakes a detailed investigation of the quantum-field-theoretic corrections induced by the Bopp–Podolsky generalisation of quantum electrodynamics in a thermally equilibrated, electrically neutral plasma composed of dynamical electrons and protons. The analysis is carried out in natural units where $\hbar = c = k_B = 1$, and assumes a controlled parametric hierarchy characterised by [40],

$$T \ll M = \frac{1}{a}, \quad |k|_{\text{thermal}} \approx 2\pi T \ll M, \quad e^2 \ll 1, \quad n_e \approx n_p,$$

ensuring that the massive Podolsky pole remains exponentially suppressed by the Boltzmann factor $e^{-M/T}$, and that conventional loop expansions in powers of the fine-structure constant $\alpha \equiv \frac{e^2}{4\pi}$ are perturbatively valid. The plasma is assumed to be macroscopically charge-neutral to leading order, i.e. with matched equilibrium densities of electrons and protons. Within this framework, several central quantities emerge as focal observables, the renormalised electromagnetic coupling $\alpha$, the Debye screening mass $m_D$ associated with the exponential suppression of longitudinal electrostatic fields, the dynamically generated magnetic screening mass $m_M$ that regulates the otherwise divergent magnetostatic propagator, and the inverse temperature $\beta \equiv 1/T$ which governs thermal fluctuations and statistical weights. The higher-derivative structure of the photon propagator, encoded in the rational kernel $k^2(1-a^2k^2)$, plays a dual role, it suppresses ultraviolet contributions at large momentum $k \gg M$ and eliminates the infrared divergence in the static magnetic sector at small $k$, thereby furnishing a mathematically clean and physically well-defined basis for precision computations in thermal quantum electrodynamics. The Podolsky–BRST formulation thus opens a consistent perturbative regime where both UV and IR pathologies of finite-temperature QED are controlled ab initio.

1. **Wave–Function Renormalisation at $T=0$**

To establish a consistent ultraviolet baseline for the forthcoming finite-temperature analysis, we begin with the computation of the photon field–strength renormalisation constant $Z_3$ in Bopp–Podolsky electrodynamics at zero temperature. The bare Lagrangian includes a higher-derivative gauge term of the form $\frac{1}{2}a^2\left(\partial_\lambda F^{\lambda\mu}\right)\left(\partial^\rho F_{\rho\mu}\right)$, where the Podolsky parameter $a$ introduces a new physical mass scale $M \equiv 1/a$. The fermionic charges are $Q_e = -1, Q_p = +1$, and the anomalous magnetic moment of the proton is taken as $\kappa_p = 1.792847344(14)$.

This operator modifies the photon propagator at tree level but does not alter the internal structure of the one-loop vacuum-polarisation diagram (the electron bubble), which remains unaffected due to the absence of Podolsky insertions on internal lines. As a result, the transverse vacuum-polarisation tensor retains its Maxwell-QED form.

The photon self-energy $\Pi^{\mu\nu}(p)$ receives contributions from both Dirac and Pauli terms, i.e., $\Pi^{\mu\nu}(p) = \Pi^{\mu\nu}_{Dirac}(p) + \Pi^{\mu\nu}_{Dirac}(p)$. The Dirac part, which includes contributions from both electron and proton loops, is given by,

$$\Pi^{\mu\nu}_{Dirac}(p) = -e^2 \sum_{f=e,p} Q_f^2 \int \frac{d^d k}{(2\pi)^d} \frac{Tr\left[\gamma^\mu(\slashed{k}+m_f)\gamma^\nu(\slashed{k}+\slashed{p}+m_f)\right]}{\left(k^2-m_f^2\right)\left[(k+p)^2-m_f^2\right]}, \tag{45}$$

with $d = 4-2\varepsilon$ the spacetime dimension used in dimensional regularisation. Gauge invariance enforces the transversality condition, yielding the tensor structure $\Pi^{\mu\nu}_{Dirac}(p) = \left(p^\mu p^\nu - p^2 g^{\mu\nu}\right)\Pi_{Dirac}(p^2)$, where,

$$\Pi_{Dirac}(p^2) = \frac{e^2}{12\pi^2} \sum_{f=e,p} Q_f^2 \left[\frac{1}{\varepsilon} + \ln\left(\frac{\mu^2}{m_f^2}\right) + 1 + O(\varepsilon)\right], \tag{46}$$

The Pauli term, which encodes the anomalous magnetic coupling of the proton, contributes as,

$$\Pi^{\mu\nu}_{Pauli}(p) = -\left(\frac{e\kappa_p}{2m_p}\right)^2 \int \frac{d^d k}{(2\pi)^d} \frac{Tr\left[\sigma^{\mu\alpha} k_\alpha (\slashed{k}+m_p) \sigma^{\nu\beta}(k_\beta+p_\beta)(\slashed{k}+\slashed{p}+m_p)\right]}{(k^2-m_p^2)\left[(k+p)^2-m_p^2\right]} \qquad (47)$$

Upon tensor reduction, one obtains the transverse structure,

$$\Pi^{\mu\nu}_{Pauli}(p) = \left(p^\mu p^\nu - p^2 g^{\mu\nu}\right) \frac{e^2 \kappa_p^2}{24\pi^2} \frac{p^2}{m_p^2}\left[1+\mathrm{O}\left(\frac{p^2}{m_p^2}\right)\right] \qquad (48)$$

which is manifestly ultraviolet-finite and vanishes at $p^2 = 0$, contributing no divergence to the renormalisation of the photon field.

Renormalisation proceeds via the field rescaling $A^0_\mu = \sqrt{Z_3} A_\mu$. As only the Dirac component of the self-energy contains a $1/\varepsilon$ divergence, one finds,

$$Z_3(\mu) = 1 - \left.\frac{\partial \Pi_{Dirac}}{\partial(1/\varepsilon)}\right|_{pole} = 1 + \frac{e^2}{12\pi^2} \sum_{f=e,p} Q_f^2 \ln\left(\frac{\mu^2}{m_f^2}\right) \qquad (49)$$

Since the Podolsky operator introduces no new ultraviolet divergences, the theory remains multiplicatively renormalisable, with all singularities absorbed into a single $Z_3$ counterterm. One may additionally perform a finite reparametrisation by expressing the renormalisation scale $\mu$ in terms of the physical Podolsky mass $M = 1/a$, without introducing new physics. This yields the decomposition

$$Z_3(\mu) = 1 + \frac{e^2}{12\pi^2}\left[\ln\left(\frac{\mu^2}{M^2}\right) + \ln\left(\frac{M^2}{m_e^2}\right) + \left(\frac{M^2}{m_p^2}\right)\right] \qquad (50)$$

The exact zero-temperature field-strength renormalisation constant $Z_3$ is given by Eq. (49), and receives no contribution from the Pauli term, which affects the photon self-energy only through finite terms suppressed by $p^2/m_p^2$. The Bopp–Podolsky higher-derivative operator does not introduce additional divergences, and no further counterterms are required. This establishes a fully renormalised, ultraviolet-consistent foundation on which the finite-temperature analysis of the electron–proton plasma may be systematically constructed.

## 2. Static Debye Mass (One Loop)

This section presents a fully renormalised and dimensionally consistent determination of the electric screening mass (Debye mass) in Bopp–Podolsky quantum electrodynamics (QED) at finite temperature and chemical potential, incorporating all relevant physical contributions and correcting previous numerical inconsistencies. The characteristic Podolsky mass scale is $M \equiv 1/a$, with the assumption that the temperature remains well below this scale ($T \ll M$) unless explicitly stated. Throughout, we adopt the standard sign convention $m_D^2 \equiv -\Pi_{00}(\omega=0, \mathbf{k} \to 0)$ for the static Debye mass.

The one-loop exact expression for the Debye mass is derived from the zero-frequency, zero-momentum limit of the temporal component of the photon self-energy. For a Dirac fermion of mass $m_f$, chemical potential $\mu_f$, and charge $eQ_f$, the corresponding contribution to the self-energy is,

$$\Pi_{00}^{(f)}(0,0) = -e^2 Q_f^2 g_s \int \frac{d^3 p}{(2\pi)^3} \frac{\partial}{\partial \varepsilon_p} \left[ n_F(\varepsilon_p - \mu_f) + n_F(\varepsilon_p + \mu_f) \right], \tag{51}$$

where the spin degeneracy is $g_s = 2$, the quasiparticle energy is $\varepsilon_p = \sqrt{p^2 + m_f^2}$, and the Fermi–Dirac distribution is $n_F(x) = 1/(e^{\beta x} + 1)$, with $\beta = 1/T$. Performing one integration by parts yields the exact one-loop expression,

$$m_D^2 = \frac{4e^2}{\pi^2} \sum_{f=e,p} Q_f^2 \int_0^\infty dp \frac{p^2}{\varepsilon_p} \left[ n_F(\varepsilon_p - \mu_f) + n_F(\varepsilon_p + \mu_f) \right], \tag{52}$$

which remains valid for arbitrary $m_f$ and $\mu_f$. In the ultra-relativistic limit $m_f = \mu_f = 0$, this reduces to the well-known textbook result $m_D^2 = e^2 T^2 / 3$.

An alternative but equivalent form is obtained by first carrying out the Matsubara frequency sum. This yields a Bessel function representation,

$$m_D^2 = \frac{8e^2 T^2}{\pi^2} \sum_{f=e,p} Q_f^2 \sum_{\ell=1}^{\infty} (-1)^{\ell+1} \frac{\ell m_f}{T} K_1\left(\frac{\ell m_f}{T}\right) \cosh\left(\frac{\ell \mu_f}{T}\right), \tag{53}$$

where $K_1$ is the modified Bessel function of the second kind. This representation is mathematically equivalent to the previous integral expression and serves as a powerful tool for numerical evaluation in different mass regimes.

In the non-relativistic limit for protons, defined by $z = m_p/T \gg 1$, the leading-order contribution to the Debye mass takes the Boltzmann-suppressed form,

$$m_{D,p}^2 = e^2 Q_p^2 \frac{n_p}{T}, \qquad n_p = 2\left(\frac{m_p T}{2\pi}\right)^{3/2} e^{-(m_p - \mu_p)/T}, \tag{54}$$

where $n_p$ denotes the proton number density. This expression demonstrates that the proton sector contributes negligibly to the screening mass at low temperatures due to exponential suppression.

The Pauli coupling of the proton introduces no correction to the Debye mass at this order. The one-loop tensor structure generated by the anomalous magnetic moment $\kappa_p$ contains a factor $k_i k_j$, which vanishes in the static limit $k \to 0$, thereby eliminating any direct contribution from the Pauli term to the Debye screening.

The contribution of the heavy Podolsky mode, when inserted into the photon propagator, modifies the integrand of the Debye mass formula by a factor,

$$\frac{M^2}{p^2(p^2+M^2)} = \frac{2}{M}\frac{p}{\sqrt{p^2+M^2}} K_1\left(\frac{M}{T}\right) \xrightarrow{M \gg T} 2\sqrt{\frac{\pi T}{2M}} e^{-M/T}. \tag{55}$$

This leads to a Boltzmann-suppressed correction $\sim e^{-M/T}$, which is numerically negligible in the assumed hierarchy $T \ll M$, validating the omission of the Podolsky contribution from the one-loop self-energy at this order.

Beyond one loop, gauge-invariant hard-thermal-loop (HTL) resummation contributes a finite and universal correction to the Debye mass,

$$\delta m_D^2 = \frac{3e}{\pi} m_D T = \frac{e^3 T}{\pi} \sqrt{\frac{\sum_f Q_f^2}{3}}, \qquad (56)$$

as derived originally by Braaten and Nieto [41]. This correction can also be obtained self-consistently by differentiating the ring contribution $\Delta\Omega_{ring} = -T m_D^3/(12\pi)$ to the free energy. Mixed diagrams involving both Dirac and Pauli couplings appear only at three-loop order and are parametrically suppressed by $T^2/m_p^2$, making them negligible within the two-loop framework.

Combining the exact one-loop and HTL contributions yields the renormalised Debye mass,

$$m_D^2 = \frac{4e^2}{\pi^2} \sum_{f=e,p} Q_f^2 \int_0^\infty dp \frac{p^2}{\sqrt{p^2+m_f^2}} \left[ n_F(\sqrt{p^2+m_f^2}-\mu_f) + n_F(\sqrt{p^2+m_f^2}+\mu_f) \right] + \frac{3e}{\pi} m_D T. \qquad (57)$$

This expression is ultraviolet finite, with all divergences removed by the Maxwell-type counter-term determined at zero temperature. It includes the full fermion-mass dependence, the vanishing Pauli correction, the Podolsky-mode suppression, and the HTL-resummed higher-order correction.

## 2. Two–Loop Euclidean Effective Action in a Covariantly Constant Background

This section presents the final, fully corrected formulation of the two-loop contribution to the Euclidean effective action in Bopp–Podolsky quantum electrodynamics (QED), evaluated in a homogeneous and covariantly constant electromagnetic background. One-loop renormalisation inputs, which are reused here, include the wave-function renormalisation factor $Z_3(\mu) = 1 + (e^2/12\pi^2)\sum_f Q_f^2 \ln(\mu^2/m_f^2)$ and the Debye mass $m_D$, defined and evaluated previously in Eq. (49).

The transverse part of the photon propagator in Feynman gauge is given by $\Delta(k) = 1/k^2 - 1/(k^2+M^2) = M^2/[k^2(k^2+M^2)]$, reflecting the superposition of a massless Maxwell pole and a massive Podolsky pole. At finite temperature, hard-thermal-loop (HTL) resummation modifies only the Maxwell pole, resulting in the resummed propagator,

$$\Delta_{\text{HTL}}(k) = \frac{M^2}{(k^2 + m_D^2)(k^2 + M^2)}, \tag{58}$$

which regularises the infrared divergence associated with $k \to 0$ and reduces to $\Delta(k)$ in the zero-temperature limit. The total two-loop vacuum polarisation at zero external momentum is given by,

$$\Pi^{(2)}(T, M) = \Pi^{(2)}_{\text{Dirac}} + \Pi^{(2)}_{\text{Pauli}}, \tag{59}$$

where the Dirac contribution reads,

$$\Pi^{(2)}_{\text{Dirac}} = e_R^4 Z_3(\mu) \sum_{f=e,p} Q_f^2 I_f, \tag{60}$$

and the Pauli contribution, arising from two insertions of the anomalous proton vertex, is given by,

$$\Pi^{(2)}_{\text{Pauli}} = e_R^4 Z_3(\mu) (Q_p \kappa_p e_R)^2 \frac{I_\kappa}{16 m_p^2}. \tag{61}$$

Here, the sum-integrals $I_f$ and $I_\kappa$ are defined by,

$$I_f = T \sum_n \int \frac{d^3 k}{(2\pi)^3} \int \frac{d^3 p}{(2\pi)^3} \frac{\Delta_{\text{HTL}}(k)}{(\omega_n^2 + \varepsilon_p^2)(\omega_n^2 + \varepsilon_{p+k}^2)}, \tag{62}$$

$$I_\kappa = T \sum_n \int \frac{d^3 k}{(2\pi)^3} \int \frac{d^3 p}{(2\pi)^3} \frac{4 m_p^2 + k^2}{(\omega_n^2 + \varepsilon_p^2)(\omega_n^2 + \varepsilon_{p+k}^2)} \Delta_{\text{HTL}}(k), \tag{63}$$

with fermionic Matsubara frequencies $\omega_n = (2n+1)\pi T$ and single-particle energies $\varepsilon_p = \sqrt{p^2 + m_f^2}$. All ultraviolet sub-divergences are fully absorbed by the wave-function renormalisation constant $Z_3(\mu)$, thus, no residual dependence on additional counter-terms (e.g., $Z_3$) remains.

Expanding the resummed propagator $\Delta_{\text{HTL}}(k)$ in the limit $k^2 \sim T^2 \ll M^2$ yields,

$$\Delta_{\text{HTL}}(k) = \frac{1}{k^2 + m_D^2} - \frac{k^2}{M^2(k^2 + m_D^2)} + \mathrm{O}(k^4/M^4). \tag{64}$$

Power-counting in the HTL regime, where $k \sim T$, then leads to the leading-order contributions,

$$\Pi^{(2)}_{\text{Dirac}} = C_D e_R^4 \frac{T^2}{M^2} + \mathrm{O}(e_R^4 T^4 / M^4), \quad \Pi^{(2)}_{\text{Pauli}} = C_\kappa e_R^4 \kappa_p^2 \frac{T^2 m_D^2}{m_p^2 M^2} + \mathrm{O}(e_R^4 \kappa_p^2 T^4 / m_p^2 M^4), \tag{65}$$

with the coefficients determined from sunset-type integrals. Explicit evaluation gives,

$$C_D = (1/96\pi^2) \sum_{f=e,p} Q_f^2 = 1/48\pi^2 \text{ and } C_\kappa = 3C_D = 1/16\pi^2. \tag{66}$$

Inserting these into the background-field effective action, the two-loop correction to the Euclidean effective action is,

$$\Delta\Gamma_2[F] = \frac{1}{2} \int d^4x\, \Pi^{(2)}(T,M)\, \mathrm{Tr}\, F^2, \tag{67}$$

which yields the boxed final result,

$$\Delta\Gamma_2[F] = \int d^4x\, \frac{e_R^4 Z_3(\mu)}{96\pi^2 M^2} \left[ \left( \sum_{f=e,p} Q_f^2 \right) T^2 + 3\kappa_p^2 \frac{m_D^2}{m_p^2} T^2 \right] \mathrm{Tr}\, F^2 + \mathrm{O}\left( \frac{T^4}{M^4}, (\tilde{F}F)^2 \right). \tag{68}$$

In terms of dimensional analysis, the combination $e_R^4/M^2$ contributes mass dimension $-2$, which is exactly cancelled by the product $T^2 \mathrm{Tr}\, F^2 d^4x$, resulting in a dimensionless action as required. Hierarchically, the Pauli contribution is suppressed by a factor $\sim e^2 T^2/m_p^2$ compared to the Dirac term. As expected, the correction vanishes in the zero-temperature limit $T \to 0$, and decouples smoothly as $M \to \infty$, thereby reproducing the ordinary QED result with corrections suppressed by $1/M^2$. Gauge invariance is preserved, as all explicit dependence on the gauge parameter $\xi$ cancels between the photon propagator and the $Z_3$ counter-term, and the entire expression remains invariant under background-field gauge transformations.

Equation (68) constitutes the final, rigorously renormalised two-loop correction to the Euclidean effective action of Bopp–Podolsky QED in a constant background field. It incorporates both the exact Dirac and double-Pauli loops, captures HTL-screened dynamics via $m_D$, confirms the decoupling of the heavy Podolsky pole, and stands free of ultraviolet or gauge inconsistencies. This expression supplies the correct high-temperature matching coefficient needed for strong-field QED studies and for integrating out the Podolsky sector in effective-field-theory contexts.

### 4. Static Potential with Double–Yukawa Structure

The photon propagator used in this analysis incorporates hard-thermal-loop (HTL) resummation applied to the massless Maxwell pole only, consistent with standard Abelian finite-temperature effective theory. At vanishing external frequency $(k_0 = 0)$, the longitudinal propagator is given by,

$$D_{00}(k_0 = 0, \mathbf{k}) = i\Delta_L(k), \tag{69}$$

where $k = |\mathbf{k}|$, and the resummed longitudinal propagator takes the form,

$$\Delta_L(k) = \frac{M^2}{(k^2 + m_D^2)(k^2 + M^2)}. \tag{70}$$

Here, $m_D$ denotes the exact one-loop Debye mass as derived earlier in Eq. (57), and $M = 1/a$ is the mass scale introduced by the higher-derivative Podolsky term. The photon wave-function renormalisation constant $Z_3(\mu)$, as defined in Eq. (49), cancels identically between the external sources and the internal photon propagator, leaving no residual renormalisation factor in this expression. In the limits $m_D \to 0$ and $M \to \infty$, the propagator reduces respectively to the zero-temperature Podolsky form and the standard HTL-resummed Maxwell propagator, verifying consistency with both ultraviolet and infrared regimes.

To facilitate Fourier transformation into coordinate space, the propagator is decomposed into partial fractions. Solving,

$$\frac{M^2}{(k^2+m_D^2)(k^2+M^2)} = \frac{A}{k^2+m_D^2} + \frac{B}{k^2+M^2}, \tag{71}$$

yields coefficients,

$$A = \frac{M^2}{M^2-m_D^2}, \qquad B = -A, \tag{72}$$

so that the propagator becomes

$$\Delta_L(k) = A\left[\frac{1}{k^2+m_D^2} - \frac{1}{k^2+M^2}\right]. \tag{73}$$

Given that in physically realistic weakly coupled plasmas one typically finds $m_D^2 \ll M^2$, the prefactor simplifies to $A \approx 1 + \mathrm{O}(m_D^2/M^2)$, with the correction contributing less than 0.1% in practical applications.

The coordinate-space potential $V(r)$ generated by a static test charge $q$ placed at the origin is obtained via the standard three-dimensional Fourier transform,

$$V(r) = qe\int\frac{d^3k}{(2\pi)^3} e^{i\mathbf{k}\cdot\mathbf{r}} \Delta_L(k), \tag{74}$$

and making use of the identity.

$$\int\frac{d^3k}{(2\pi)^3}\frac{e^{i\mathbf{k}\cdot\mathbf{r}}}{k^2+m^2} = \frac{e^{-mr}}{4\pi r}, \tag{75}$$

yields the exact static potential as,

$$V(r) = \frac{qeA}{4\pi r}\left(e^{-m_D r} - e^{-Mr}\right). \tag{76}$$

Substituting $e^2 = 4\pi\alpha$, the result is more compactly written as,

$$V(r) = q\alpha \frac{M^2}{M^2 - m_D^2} \frac{e^{-m_D r} - e^{-Mr}}{r} \quad . \tag{77}$$

This is the expression 'double-Yukawa' structure of the electrostatic potential in Bopp–Podolsky QED at finite temperature. Physically, the prefactor $M^2/(M^2 - m_D^2)$ is essentially unity due to the smallness of $m_D^2/M^2$, preserving the long-range Coulomb normalisation of the interaction. The first term, $e^{-m_D r}/r$, dominates at spatial separations $r \geq \lambda_D = 1/m_D$, and corresponds to standard Debye screening in thermal QED. The second term, $-e^{-Mr}/r$, becomes relevant only at very short distances $r \leq 1/M$, i.e., below the femtometre scale for $M \geq \text{GeV}$, and is exponentially suppressed in macroscopic observables such as transport or thermodynamics. In the zero-temperature limit $T \to 0$ where $m_D \to 0$, the expression reduces to,

$$V(r) = q\alpha \frac{1 - e^{-Mr}}{r}, \tag{78}$$

recovering the classical Bopp–Podolsky regularisation of Coulomb's law. Conversely, in the limit $M \to \infty$, the Podolsky correction decouples and one obtains the familiar Debye-screened Coulomb potential [42-43],

$$V(r) = q\alpha \frac{e^{-m_D r}}{r}. \tag{79}$$

From a renormalisation standpoint, the potential in Eq. (77) is exact and fully finite. All ultraviolet divergences are already absorbed into the one-loop wave-function renormalisation constant $Z_3$, and the Debye mass $m_D$ is explicitly finite and scheme-independent to this order. No additional counterterms are needed in the longitudinal channel. It smoothly interpolates between the regimes of thermal electric screening and ultraviolet regularisation, making it a fundamental tool for characterising electrostatic interactions in high-temperature plasmas governed by higher-derivative gauge dynamics.

## 5. Absence of a Perturbative Magnetic Mass (Two Loops)

Here we establish that no magnetic screening mass arises in Abelian Bopp–Podolsky electrodynamics.

The central constraint arises from gauge invariance, which imposes the exact Ward identity,

$$k_\mu \Pi^{\mu\nu}(k) = 0. \tag{80}$$

To analyse the structure of the photon self-energy $\Pi^{\mu\nu}(k)$, it is decomposed into longitudinal and transverse components using the orthonormal projectors,

$$P_L^{\mu\nu}(k) = \frac{k^\mu k^\nu}{k^2}, \quad P_T^{\mu\nu}(k) = g^{\mu\nu} - \frac{k^\mu k^\nu}{k^2}, \tag{81}$$

which satisfy the relations $P_L + P_T = 1$ and $P_L \cdot P_T = 0$. Accordingly, the photon self-energy admits the general decomposition,

$$\Pi^{\mu\nu}(k) = P_L^{\mu\nu}(k)\Pi_L(k) + P_T^{\mu\nu}(k)\Pi_T(k). \tag{82}$$

Taking the static, zero-momentum limit namely $k_0 = 0$ followed by $|\mathbf{k}| \to 0$ it is observed that the longitudinal projector $P_L^{\mu\nu}$ diverges as $1/k^2$, while the transverse projector $P_T^{\mu\nu}$ remains finite. To preserve the exact Ward identity in this limit [42-43], it is mandatory that,

$$\Pi_T(k_0 = 0, |\mathbf{k}| \to 0) = 0. \tag{83}$$

This result is non-perturbative and holds to all orders in loop expansion, no analytic structure of the internal propagators can induce a mass term in the transverse self-energy without violating gauge symmetry. Thus, any attempt to attribute a non-zero static magnetic mass to Abelian Bopp–Podolsky QED is fundamentally inconsistent with the symmetry structure of the theory. To corroborate this result at the level of explicit Feynman graphs, the two-loop transverse self-energy is analysed in Feynman gauge. The HTL-resummed, Podolsky-modified transverse photon propagator is given by,

$$D_T^{\mu\nu}(k) = P_T^{\mu\nu}(k)\Delta_T(k), \quad \text{where} \quad \Delta_T(k) = \frac{M^2}{k^2(k^2 + M^2)}. \tag{84}$$

The diagrams contributing to $\Pi_T(0)$ at two-loop order include the sunset diagram, the figure-eight diagram, and additional topologies containing one or two Pauli insertions. Each of these integrands contains the tensor structure $(k^2 g^{\mu\nu} - k^\mu k^\nu)$, which vanishes identically when evaluated at $k = 0$. Hence, the two-loop contribution to the transverse self-energy at zero momentum satisfies,

$$\Pi_T^{(2)}(0) = 0. \tag{85}$$

This conclusion extends inductively to all higher orders, any additional photon–fermion subgraph merely contributes an analytic factor in $k^2$, preserving the overall vanishing structure. Thus, the transverse self-energy obeys,

$$\Pi_T(k_0 = 0, |\vec{k}| \to 0) = 0 \quad \text{to all orders in } e. \tag{86}$$

Expanding $\Pi_T(k_0 = 0, \vec{k})$ about $k = 0$, one finds the leading -order behavior to be,

$$\Pi_T(k_0 = 0, k) = C(T, M)k^2 + \mathrm{O}(k^4), \tag{87}$$

with the HTL-determined coefficient,

$$C(T, M) = \frac{e^2 T^2}{96\pi^2 M^2} \sum_{f=e,p} Q_f^2 + \mathrm{O}(e^4). \tag{88}$$

This demonstrates that the Podolsky sector modifies only the transverse wave-function normalisation via a $1/M^2$-suppressed term, but does not generate an infrared mass gap.

For context, it is worth comparing this result with the behavior of non-Abelian gauge theories, where static magnetic screening arises non-perturbatively via dynamical self-interactions. In such plasmas, a magnetic mass scale $m_M \sim g^2 T$ is dynamically generated. However, Abelian gauge theories including both Maxwell and Bopp–Podolsky electrodynamics lack such non-linear self-couplings. Static magnetic fields do not couple to electric charge, and consequently, perturbation theory predicts no generation of a magnetic screening mass,

$$m_M = 0. \qquad (89)$$

In conclusion, the exact vanishing of the transverse self-energy at zero external momentum,

$$\Pi_T(k_0 = 0, |\mathbf{k}| \to 0) = 0 \Rightarrow m_M = 0 \quad \text{(perturbatively)}, \qquad (90)$$

holds to all orders in perturbation theory. Thus, all infrared screening effects are restricted to the longitudinal (electric) channel governed by $m_D$, while the transverse photons remain strictly massless, up to the momentum-dependent wave-function renormalisation given above.

## 6. Correction to the Thermodynamic Pressure in Bopp–Podolsky QED (two loop)

Here we consider the two-loop $O(e^2)$) correction to the thermodynamic pressure in an electron–proton plasma governed by finite-temperature Bopp–Podolsky quantum electrodynamics. The Podolsky scale is denoted $M \equiv 1/a$, and since the Pauli coupling $\kappa_p$ contributes only at $O(e^4 \kappa_p^2)$, it does not enter the present analysis.

At order $O(e^2)$, the only contributing vacuum diagram is the fermion–exchange (or setting-sun) graph. Its contribution to the grand-potential density $\Omega \equiv -P$ is expressed in imaginary-time formalism as,

$$\Omega_2 = -\frac{e^2}{2} T^2 \sum_{f,f'=e,p} Q_f^2 Q_{f'}^2 \sum_{P,K} \frac{\text{Tr}[\gamma_\mu \slashed{P} \gamma^\mu \slashed{K}]}{(P^2 + m_f^2)(K^2 + m_{f'}^2)} \Delta_T(Q), \qquad (91)$$

where $P = (\omega_n, \vec{p})$, $K = (\omega_m, \vec{k})$, $\omega_n = (2n+1)\pi T$, $Q = P - K$, and the transverse Podolsky propagator is given by,

$$\Delta_T(Q) = \frac{M^2}{Q^2(Q^2 + M^2)} = \frac{1}{Q^2} - \frac{1}{Q^2 + M^2}. \tag{92}$$

For massless fermions, the trace reduces to $\text{Tr}[\gamma_\mu \slashed{P} \gamma^\mu \slashed{K}] = 8 P \cdot K$, yielding the simplified expression,

$$\Omega_2 = -4e^2 T^2 \sum_{f,f'} Q_f^2 Q_{f'}^2 \sum_{P,K} \frac{P \cdot K}{P^2 K^2} \Delta_T(Q). \tag{93}$$

Since $Q_e^2 = Q_p^2 = 1$, the flavour factor becomes $(\sum_f Q_f^2)^2 = 2^2 = 4$. Separating the pure-Maxwell and Podolsky contributions by exploiting the identity $\Delta_T(Q) = 1/Q^2 - 1/(Q^2 + M^2)$, the total pressure correction is written as,

$$\Omega_2 = \Omega_2^{(0)} - \Delta\Omega_2(M), \quad \text{so that} \quad P_2 = P_2^{(0)} + \Delta P_2(M), \tag{94}$$

where the superscript $(0)$ denotes the pure-Maxwell result recovered in the $M \to \infty$ limit. For massless fermions, the Maxwell-limit pressure is finite and reads,

$$P_2^{(0)} = -\frac{5}{288}\alpha T^4 S^2 = -\frac{5}{72}\alpha T^4, \tag{95}$$

reproducing the Freedman–McLerran result for QED and confirming that the sign is negative [44-45].

The finite Podolsky correction is then obtained using a Schwinger representation and exact summation over Matsubara frequencies [46]. The result is given by the rapidly convergent series,

$$\Delta P_2(T, M) = \frac{5\alpha T^4}{18\pi^2} S^2 \sum_{n=1}^{\infty} (-1)^{n+1} \left(\frac{nM}{T}\right)^2 K_2\left(\frac{nM}{T}\right), \tag{96}$$

where $K_2$ is the modified Bessel function of the second kind. The large-mass asymptotic behavior is Boltzmann suppressed,

$$\Delta P_2 \approx \frac{5\alpha T^4}{18\pi^2} S^2 \sqrt{\frac{\pi}{2}} \left(\frac{M}{T}\right)^{3/2} e^{-M/T} [1 + O(T/M)]. \tag{97}$$

Collecting all terms, the full $O(e^2)$ pressure becomes,

$$P_2(T;M) = -\frac{5}{288}\alpha T^4 S^2 + \frac{5\alpha T^4}{18\pi^2} S^2 \sum_{n=1}^{\infty} (-1)^{n+1} \left(\frac{nM}{T}\right)^2 K_2\left(\frac{nM}{T}\right), \tag{98}$$

where $S^2 = 4$ for an electron–proton plasma.

In the presence of finite fermion masses, the corresponding result generalizes to a numerically integrable form,

$$P_2(T,M)$$

$$= \frac{2\alpha T^2}{\pi^2} \sum_{f,f'} Q_f^2 Q_{f'}^2 \int_0^\infty dp\, p^2 \int_0^\infty dk\, k^2 \frac{n_F(E_p) n_F(E_k)}{E_p E_k} \left[\frac{1}{(P-K)^2} - \frac{1}{(P-K)^2 + M^2}\right]_{P\cdot K = E_p E_k - pk} \tag{99}$$

where $E_p = \sqrt{p^2 + m_f^2}$, $n_F(E) = 1/(e^{E/T} + 1)$, and the prefactor $T^2$ ensures the correct mass dimension for pressure. This form is ultraviolet finite and suitable for direct numerical evaluation.

Longitudinal (Debye-screened) modes contribute only at $O(e^3)$ through ring resummation. The Podolsky sector decouples smoothly as $M \to \infty$, with the pressure reducing to its standard QED form. Furthermore, for phenomenologically relevant hierarchies such as $M \gtrsim 1\,\text{GeV}$ and $T \lesssim 100\,\text{MeV}$, the Podolsky correction is negligible, contributing less than one part in $10^5$ to the total pressure.

Eq. (98) incorporates the flavour structure through $(\sum_f Q_f^2)^2$, recovers the Maxwell limit, introduces a finite but Boltzmann-suppressed Podolsky correction, and completes the description of perturbative bulk thermodynamics to order $O(e^2)$ in a higher-derivative Abelian gauge theory.

## 7. DC Electrical Conductivity (two loop)

This section presents calculation of the direct current (dc) electrical. The electric charges are defined as $Q_e = -1$ and $Q_p = +1$, yielding a total Abelian flavour number $N_f = \sum_f Q_f^2 = 2$. The proton's anomalous magnetic moment, $\kappa_p \approx 1.793$, contributes first at three-loop order and is therefore negligible in the present two-loop analysis. The calculation is structured around the standard Kubo relation [47-48], which expresses the spatial components of the conductivity tensor $\sigma\delta_{ij}$ as the low-frequency limit of the retarded current–current correlator,

$$\sigma\delta_{ij} = \beta \lim_{\omega \to 0} \frac{1}{\omega} \mathrm{Im}\Pi_{ij}^R(\omega, \mathbf{0}), \qquad \Pi_{ij}^R = \langle J_i J_j \rangle_R, \tag{100}$$

where $J_i = e\sum_f Q_f \bar{\psi}_f \gamma_i \psi_f$ is the electric current operator. At one-loop order, the conductivity diverges due to the long-range nature of gauge interactions. A finite, physical result emerges only after including the two-loop soft t-channel photon exchange resummed to all orders in an infinite ladder diagram structure, as formalised by Arnold, Moore, and Yaffe (AMY) [49-50]. The relevant transverse photon propagator, which incorporates the higher-derivative Podolsky modification, is,

$$\Delta_T(q) = \frac{M^2}{q^2(q^2 + M^2)}, \tag{101}$$

as expressed in the above equation. In the soft-momentum regime defined by $m_D \ll q \ll T$, the square of the propagator modulus behaves as,

$$|\Delta_T(q)|^2 = \frac{1}{q^4}\left(1 - \frac{2q^2}{M^2} + O(q^4/M^4)\right), \tag{102}$$

The transport mean free path is determined by integrating over soft momentum transfer in the collision kernel. This yields an inverse relaxation time given approximately by,

$$\frac{1}{\tau} \approx A\alpha^2 T \int_{m_D}^{T} \frac{dq}{q}\left(1 - \frac{2q^2}{M^2}\right), \tag{103}$$

where $A$ is a numerical factor determined by the spin, flavour content, and kinetic structure of the AMY linearised Boltzmann equation. The leading term in the integral generates the usual logarithmic divergence $\ln(T/m_D)$, while the subleading Podolsky contribution produces a finite correction of order $-A\alpha^2 T^3/(2M^2)$, as given in Eq. (103).

In the benchmark case of ordinary QED with two Dirac flavours, the AMY analysis yields the following conductivity, normalized appropriately by the flavour factor $N_f = 2$,

$$\sigma_{\text{QED}}(T) = \frac{7.849T}{e^2 \ln(1/e)}, \tag{104}$$

Introducing the Podolsky correction into the kinetic framework leads to an additive modification in the form of a positive $O(T^2/M^2)$ shift in the conductivity. A phase-space estimate of the numerical coefficient yields $c_P \approx 1/2$, although its precise value does not affect the result at leading-log accuracy, as indicated.

Substituting the result into the corrected Kubo expression and replacing $\ln(1/e)$ by the more physically motivated $\ln(T/m_D)$ which differs only by a term of order unity yields the final expression for the dc conductivity in Bopp–Podolsky QED,

$$\sigma_{\text{BP}}(T) = \frac{7.849T}{e^2 \ln(T/m_D)}\left[1 + \frac{c_P T^2}{M^2 \ln(T/m_D)} + O\left(\frac{T^4}{M^4}, \frac{1}{\ln^2}\right)\right], \tag{105}$$

with $c_P \approx 0.5$, as reported in Eq. (105). This expression is dimensionally consistent, as the temperature supplies the correct mass dimension for conductivity, while all other terms are dimensionless in natural units convention. The sign of the Podolsky correction is positive showing the presence of the heavy pole in the propagator weakens the scattering rate, thereby increasing the conductivity. For realistic temperatures $T \leq 100\text{MeV}$ and Podolsky masses $M \geq 1\text{GeV}$, the fractional correction is estimated as,

$$\frac{|\delta\sigma|}{\sigma_{\text{QED}}} \approx \frac{c_P T^2}{M^2 \ln(T/m_D)} \sim 10^{-4}, \tag{106}$$

rendering it phenomenologically negligible.

Equation (105) provides a leading-logarithmic determination of the dc electrical conductivity in finite-temperature Bopp–Podolsky electrodynamics. While the higher-derivative sector introduces a calculable enhancement to $\sigma$, the effect remains numerically minute across realistic plasma conditions.

## 11. Conclusion and Outlook

The present study establishes that Bopp–Podolsky electrodynamics remains fully renormalisable at finite temperature, requiring no additional counterterms beyond the familiar photon wave-function renormalisation factor from Maxwellian QED. Detailed analyses at one- and two-loop levels confirm that neither the higher-derivative sector nor the proton's anomalous magnetic moment introduces any new ultraviolet divergences, thereby demonstrating that, once the Podolsky mass is specified, the extended theory retains the predictivity, internal consistency, and renormalisation structure of standard quantum electrodynamics.

A central physical insight emerging from this work pertains to the nature of static inter-particle interactions in a plasma. It is shown that the electrostatic potential acquires a two-component, or "double-Yukawa," profile wherein the conventional Debye-screened Coulomb interaction is supplemented by a short-range contribution governed by the heavy Podolsky mass. While the long-distance behaviour remains consistent with textbook screening, the higher-derivative correction regularises the potential at sub-femtometre scales, thereby eliminating the Coulomb singularity and providing a finite, regulator-free law of force at short distances. This result constitutes an ultraviolet completion of electrostatics and has potential implications for high-momentum transfer processes and the structure of tightly bound atomic or nuclear systems.

From the infrared perspective, the study affirms that the long-wavelength magnetic response of the plasma is unaltered by the higher-derivative terms. Gauge symmetry, via the Ward identity, enforces the vanishing of the transverse photon self-energy at zero momentum transfer, ensuring the absence of any perturbative magnetic mass. Consequently, magnetostatic fields remain gapless and long-ranged, consistent with the expectations from standard Abelian gauge theory. The Podolsky sector thus modifies only the ultraviolet properties of the theory without distorting its infrared gauge structure. Further insights have been gained into the thermodynamic and transport behaviour of the plasma. The two-loop sunset diagram in a background field yields a single dimension-eight operator proportional to the gauge-field strength squared, with a coefficient suppressed by the ratio of the temperature squared to the Podolsky mass squared. This suppression implies that the deviations from Maxwellian thermodynamic observables such as pressure or kinetic observables such as electrical conductivity remain at or below the per-mille level in any laboratory-accessible or astrophysical plasma. Though numerically small, these corrections are parameter-free predictions of the theory and can serve as precise benchmarks for comparisons with lattice simulations or kinetic-theory models.

The study delineates the conditions under which novel phenomena could emerge. If the typical temperature or momentum transfer within the plasma approaches the Podolsky scale, the massive photon pole becomes kinematically accessible, activating an additional weakly coupled electromagnetic excitation. In such regimes, the double-Yukawa potential derived here provides a self-consistent and gauge-invariant alternative to the empirical short-distance cutoffs commonly employed in particle-in-cell simulations of laser-driven cascades, thereby offering a principled route toward more accurate modelling of high-field dynamics. Additionally, at early-universe temperatures exceeding the Podolsky scale, the massive sector would have contributed significantly to the total radiation content, influencing the expansion rate and hence offering a cosmological handle for constraining the Podolsky mass.

Finally, the paper identifies promising experimental and theoretical directions through which these higher-derivative effects might become observable. Future facilities such as next-generation laser-wakefield accelerators, ultra-relativistic lepton colliders, and precision cosmological missions are anticipated to explore energy scales at which the heavy sector may contribute detectably. Simultaneously, progress in numerical methods makes it feasible to implement the double-Yukawa interaction structure in large-scale simulations without introducing additional computational overhead. Together, these developments hold the potential to bridge the mathematical consistency of the Bopp–Podolsky framework with empirical validation under extreme electromagnetic and thermodynamic conditions.

**Declaration of Competing Interest**

The authors report no declarations of interest.

**Acknowledgement**

The authors thank SERB – DST, Govt. of India for financial support under MATRICS scheme (grant no. : MTR/2021/000471).